\documentclass[11pt,final]{article}

\usepackage{graphicx}
\usepackage{hyperref}
\usepackage[affil-it]{authblk}

\usepackage{amsmath,amssymb,amsfonts}

\def\GRG{Gen. Relat. Gravit.} %General Relativity and Gravitation

\def\CQG{Classical Quant. Grav.}

\def\JMP{J. Math. Phys.}

\def\LRR{Living Rev. Relat.}

\def\prd{Phys. Rev. D}

\def\RMP{Rev. Mod. Phys.}

\newcommand{\trho}{\rho}

\newcommand{\e}[1]{{\mathrm e}^{#1}}

\begin{document}

\title{\bf Source integrals of asymptotic multipole moments}

\author{{\bf Norman G{\"u}rlebeck}\thanks{\texttt{norman.guerlebeck@zarm.uni-bremen.de}}}

\affil{ZARM, University of Bremen, Am Fallturm, 28359 Bremen,
Germany, EU\\
Institute of Theoretical Physics, Charles University, V
Hole\v{s}ovi\v{c}k\'ach 2, 180 00 Praha, Czech Republic, EU
}
\date{}
\maketitle

\begin{abstract}
We derive source integrals for multipole moments that describe the behaviour of
static and axially symmetric spacetimes close to spatial infinity. We assume
isolated non-singular sources but will not restrict the matter content
otherwise. Some future applications of these source integrals of the asymptotic
multipole moments are outlined as well.
\end{abstract}

\section{Introduction}

In experiments that measure general relativistic effects, some parameters
characterizing the spacetime have to be determined. The multipole moments are
one set of such parameters. They are measured in the exterior region of
astrophysical objects like neutron stars or galaxies but also planets and
describe the gravitational field near spatial infinity. They will be called here
\emph{asymptotic multipole moments} (AMM). The bending of light and the
gravitational lensing proofed particularly useful for their measurement, see,
e.g., \cite{Wambsganss_1998,Perlick_2004,Kopeikin_2007} and references therein.
But what information can be gathered about the matter distribution and the
metric in its interior by their measurement? What does it mean to measure a
quadrupole moment of a certain amount? In Newtonian theory, the answers are
provided by the source integrals of the AMM. These integrals determine the
asymptotic multipole moments by an integration over the mass density. In general
relativity, similar expressions for the AMM are only known in approximations to
general relativity, see e.g. \cite{Thorne_1980}. Here we will present source
integrals of the AMM for static and axially symmetric spacetimes in full general
relativity. At the same time these provide quasi-local definitions of \emph{all}
asymptotic multipole moments.

\section{Preliminaries}

In this section, we shortly review several concepts necessary in the
derivation of the source integrals. We use throughout this article geometric
units $G=c=1$ and the signature of the metric is $(-,+,+,+)$.

\subsection{The line element and the field equations}\label{sec:line_element}

We concentrate on axially symmetric and static spacetimes of the Weyl form,
i.e.,
\begin{align}\label{eq:Weyl_line_element}
  ds^2=\mathrm{e}^{2 k-2U}\left(d\rho^2+d\zeta^2\right)+
  W^2\mathrm{e}^{-2U} d\varphi^2-\mathrm{e}^{2U}d t^2.
\end{align}
We do not restrict the type of matter except in that the line element
\eqref{eq:Weyl_line_element} can be introduced, see \cite{Stephani_2003}. The
metric functions $\e{2U}$ and $W$ can be expressed by the timelike Killing
vector $\xi^\alpha=(\partial_t)^\alpha$ and the Killing vector of the axial
symmetry $\eta^\alpha=(\partial_\varphi)^\alpha$
\begin{align}\label{eq:Killing_Vectors}
  \mathrm{e}^{2U}=-\xi_\alpha\xi^\alpha,\quad W^2=-\eta_\alpha\eta^\alpha
  \xi_\beta\xi^\beta.
\end{align}

Let us choose a sphere $\mathcal S_0$ of finite radius $r=R_0$ ($\rho=r
\sin\theta,~\zeta=r\cos\theta$) that covers the entire matter distribution, cf.
Fig. \ref{Fig:Curves}. Outside of $\mathcal S_0$, canonical Weyl coordinates
($W=\rho$) are introduced by virtue of one of the vacuum field equations. This
allows still a shift in the $\zeta-$coordinate, which enables us later to move
the origin with respect to which the AMM are measured. The vacuum field
equations read in canonical Weyl coordinates
\begin{align}\label{eq:field_equation_vacuum}
\begin{split}
  \Delta U=0,\quad
  k_{,\zeta}=2\rho U_{,\rho}U_{,\zeta},\quad
  k_{,\rho}=\rho\left((U_{,\rho})^2-(U_{,\zeta})^2\right),
\end{split}
\end{align}
where $\Delta=\left(\tfrac{\partial^2}{\partial\trho^2}+
\tfrac{1}{\rho}\tfrac{\partial}{\partial\rho}+
\tfrac{\partial^2}{\partial\zeta^2}\right)$.
The function $k$ is determined via a line integration, cf. the last two
equations of \eqref{eq:field_equation_vacuum}, once $U$ is known. Hence, only a
Laplace equation for $U$ remains to be solved in practise.

\subsection{The physical setting}
\begin{figure}
\begin{center}
\includegraphics[scale=0.3]{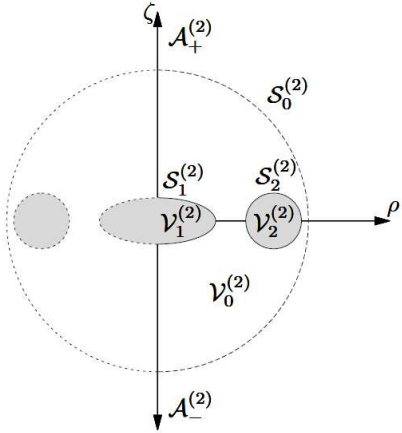}
\end{center}
%\vspace{-3.1cm}
\caption{\label{Fig:Curves}An example of the physical situations discussed here:
The surfaces of the individual matter components are denoted by $\mathcal
S^{(2)}_i=\partial \mathcal V^{(2)}_i$ with $i\geq 1$ and their respective
volumes by $\mathcal V^{(2)}_i$. The surface $\mathcal S_0^{(2)}$ describes a
circle with sufficiently large but finite radius enclosing all matter
components, cf. Sec. \ref{sec:line_element}.}
\vspace{-0.5cm}
\end{figure}
We depict in Fig. \ref{Fig:Curves} an example of a physical situation that will
be covered by the subsequent considerations. The relevant surfaces and volumes
are defined there as well. For simplicity, we allow only non-singular sources.
However, such can be incorporated into the formalism as we showed in
\cite{Gurlebeck_2012}. The 3-dimensional projection of the matter region into
an hypersurface of constant Killing-time $t$ is obtained by an rotation around the $\zeta-$axis in Fig. \ref{Fig:Curves}. In this way, the quantities $\mathcal
A_\pm^{(3)}$, $\mathcal S^{(3)}_i$ and $\mathcal V^{(3)}_i$ are defined starting
from $\mathcal \mathcal A_{\pm}^{(2)}$, $\mathcal S^{(2)}_i$ and $\mathcal
V^{(2)}_i$, respectively.

\subsection{The multipole moments}

For asymptotically flat and static spacetimes a geometric definition of AMM was
given by Geroch in \cite{Geroch_1970}. This definition was generalized and
applied by many authors, see the reviews \cite{Thorne_1980,Quevedo_1990} and references therein.
In the axially symmetric case with the line element
\eqref{eq:Weyl_line_element}, Geroch's multipole moments $M_r$ can be obtained
by an expansion of $U$ along the symmetry axis in $|\zeta|^{-1}$ :
\begin{align}
U(\rho=0,\zeta)=\sum\limits^{\infty}_{r=0}U^{(r)}|\zeta|^{-r-1}.
\end{align}
The $M_r$ follow uniquely from Weyl's multipole moments $U^{(r)}$ and vice
versa as was shown in \cite{Fodor_1989}. Therefore, we consider only the
$U^{(r)}$ here.

\subsection{The inverse scattering technique\label{sec:linear_problem_Laplace}}

Lastly, we shortly review the inverse scattering technique (IST), see e.g.
\cite{Neugebauer_2003} for a recent account. Even though the Laplace equation is
linear and the use of the IST seems artificial, the IST proves nonetheless
beneficial, because it is easily generalizable to the non-linear case of the
Ernst equation. This equation is of special interest in relativistic
astrophysics, since it describes the exterior of rotating stars. The starting
point of the IST in the present setting, i.e. the linear problem of the Laplace
equation, is given by
\begin{align}\label{eq:linear_problem_Laplace}
  \sigma_{,z}=(1+\lambda)U_{,z} \sigma, \quad
  \sigma_{,\bar z}=\left(1+\frac{1}{\lambda}\right)U_{,\bar z} \sigma,
\end{align}
where $z=\rho+{\mathrm i} \zeta$, the spectral parameter
$\lambda=\sqrt{\tfrac{K-\mathrm{i}\bar z}{K+\mathrm{i}z}}$, $K\in \mathbb C$ and
a bar denotes complex conjugation. The complex valued function $\sigma$ depends
on $z,~\bar z$ and $\lambda$. The integrability condition of Eqs.
\eqref{eq:linear_problem_Laplace} is the Laplace equation for $U$.
Therefore, having a solution $\sigma$ of Eq. \eqref{eq:linear_problem_Laplace}
yields also a solution $U$ of the Laplace equation and vice versa. The main
technical steps of the IST as described in \cite{Neugebauer_2003} are to
integrate \eqref{eq:linear_problem_Laplace} along $\mathcal A^{(2)}_\pm$, along
a circle with sufficiently large radius and along a compact curve connecting
$\mathcal A^{(2)}_+$ with $\mathcal A^{(2)}_-$.  This scheme can be carried out
partially and we quote only the results (simplified to the static case), which
are relevant for us, from \cite{Neugebauer_2003}:
\begin{subequations}\label{eq:axis_values_sigma}
\begin{align}
\begin{split}
(0,\zeta)\in \mathcal A^+:~\sigma\left(\lambda=+1,\rho=0,\zeta\right)
&=F(K)\mathrm{e}^{2U(\rho=0,\zeta)},\\
\sigma\left(\lambda=-1,\rho=0,\zeta\right)&=1,\\
(0,\zeta)\in \mathcal A^-:~\sigma\left(\lambda=+1,\rho=0,\zeta\right)
&=\mathrm{e}^{2U(\rho=0,\zeta)},\\
\sigma\left(\lambda=-1,\rho=0,\zeta\right)&=F(K).
\end{split}
\end{align}
The function $F:\mathbb C\to \mathbb C$ is given for $K\in \mathbb
R$ with $(\rho=0,\zeta=K)\in { \mathcal A^{\pm}}$ by
\begin{align}
  F(K)=
  \begin{cases}
\mathrm{e}^{-2U(\rho=0,\zeta=K)} & (0,K)\in {\mathcal A}^+\\
\mathrm{e}^{2U(\rho=0,\zeta=K)} & (0,K)\in {\mathcal A}^-
\end{cases}.
\end{align}
\end{subequations}
The integration along $\mathcal S^{(2)}_0$ does not enter the derivation of
these formulas and it forms the crucial part of our considerations in the next
section.

\section{Source integrals of Weyl's multipole moments
\label{sec:source_integrals}}

The derivation of the source integrals consists of several steps. First, the AMM
are expressed as line integrals along $\mathcal S_0^{(2)}$. This is the most
important step, because it makes it possible to determine the AMM quasi-locally.
Then these integrals will be rewritten in a coordinate independent form as
surface integrals over $\mathcal S_0^{(3)}$ by virtue of the axial symmetry.
Subsequently, Stokes' theorem is used to rewrite these as volume integrals over
$\mathcal V^{(0)}$. In the final step, it is shown that the contributions in the
vacuum regions vanish. Thus, the steps from before can be retraced to obtain the
contributions in source integral form of each individual matter component. We
suppress the details of these derivations and show only the crucial steps.

The linear problem \eqref{eq:linear_problem_Laplace} is well-defined along
$\mathcal S^{(2)}_0$ and reads:
\begin{align}\label{eq:Linear_Proble_Surface}
  \sigma_{,s}=\left[U_{,s}+\frac
  12\left(\left(\frac{1}{\lambda}+\lambda\right)U_{,s}+{\mathrm i}
  \left(\frac{1}{\lambda}-\lambda\right)U_{,n}\right)\right]\sigma,
\end{align}
where $U_{,s}$ and $U_{,n}$ are the tangential and the (outward pointing) normal
derivative of $U$ along $\mathcal S_0^{(2)}$ with respect to a parametrisation
$[s_N,s_S]\to \mathcal S_0^{(2)}$. The indices $N$ and $S$ refer to the values
of a parameter or function at the ``north'' and ``south'' pole of $\mathcal
S^{(2)}_0$, i.e., to the intersection points $(\rho=0,\zeta=\zeta_{N/S})$ of
$\mathcal S_0^{(2)}$ and the symmetry axis. Eq.
\eqref{eq:Linear_Proble_Surface} is easily integrated using the boundary values
from Eq. \eqref{eq:axis_values_sigma}:
\begin{align}\label{eq:Compatibility_Condition}
\begin{split}
 U(0,K)=&\frac{U_N-U_S}{2}+\frac{1}{4}\int\limits_{s_N}^{s_S}
 \left(\left(\lambda^{-1} +\lambda\right)U_{,s} +\mathrm
 i\left(\lambda^{-1}- \lambda\right)U_{,n}\right){\mathrm d} s.
\end{split}
\end{align}
If we expand this equation in $|K|^{-1}$, we obtain expression for Weyl's
multipole moments in terms of a line integration. Let us introduce the
abbreviations $N^{(r)}_{+}$ and $N^{(r)}_{-}$ for the expansion
coefficients $(\lambda^{-1}+ \lambda)^{(r)}$ and ${\mathrm
i}(\lambda^{-1} - \lambda)^{(r)}$ to order $r+1$, respectively. After a lengthy
but straightforward calculation they evaluate to
\begin{align}\label{eq:expansion_coefficients}
\begin{split}
  N_{-}^{(r)}&=
  \sum\limits_{k=0}^{\left\lfloor\frac{r}{2}\right\rfloor}\frac{2(-1)^{k+1}
  r!\rho^{2k+1}\zeta^{r-2k}}{4^k (k!)^2(r-2k)!},\\
  N_{+}^{(r)}&=
  \sum\limits_{k=0}^{\left\lfloor\frac{r-1}{2}\right\rfloor}\frac{2
  (-1)^{k+1}r!\rho^{2k+2}\zeta^{r-2k-1}}{4^k (k!)^2(r-2k-1)!(2k+2)}.
\end{split}
\end{align}

The $r=-1$ order in $|K|^{-1}$ of Eq. \eqref{eq:Compatibility_Condition} is
satisfied trivially and will not be considered subsequently. The orders $r\geq
0$ of Eq.
\eqref{eq:Compatibility_Condition} yield the desired quasi-local definitions of
Weyl's multipole moments:
\begin{align}\label{eq:multipole_moments}
  U^{(r)}=\frac{1}{4}\int\limits_{\mathcal S^{(2)}_0}\left(N^{(r)}_+
  U_{,\hat s}+ N^{(r)}_{-} U_{,\hat n}\right)\mathrm d \mathcal S^{(2)}_0,
\end{align}
where $U_{,\hat s}$ and $U_{,\hat n}$ are the tangential and normal derivatives
along $\mathcal S_0^{(2)}$ with respect to the unit tangent vector and the unit
normal vector, which are defined with the induced metric on $\mathcal
S^{(2)}_0$. ${\mathrm d} \mathcal S^{(2)}_0$ denotes the proper distance along
$\mathcal S^{(2)}_0$. The functions $N_{\pm}^{(r)}$ and $U$ are to be read as
functions along $\mathcal S^{(2)}_0$, i.e. as functions of $(\rho(s),\zeta(s))$.

To make the coordinate independence apparent, we express $\rho$ and $\zeta$
by scalars build from the Killing vectors. Firstly, observe that Eq.
\eqref{eq:Killing_Vectors} holds everywhere and that in vacuum we have $W=\rho$.
Additionally, the 1-form
\begin{align}
Z_\alpha=\epsilon_{\alpha\beta\gamma\delta}W^{,\beta}W^{-1}\eta^\gamma\xi^\delta
\end{align}
is well-defined and hypersurface orthogonal everywhere as well as exact in
the vacuum region. Hence, there exist a potential $Z$ and an integrating factor
$X$ such that $Z_{,\alpha}=X Z_{\alpha}$, where $X=1$ in the exterior of $\mathcal S^{(3)}_0$. In the vacuum region and in canonical Weyl coordinates, we find
$Z=\zeta+\mathrm{const}$. Since we can shift the $\zeta$-coordinate freely, we
can drop the constant of integration, which specifies the origin with respect
to which the AMM are measured. Thus, $W$ and $Z$ coincide with
$\rho$ and $\zeta$ in the vacuum region and can be used as their continuation
into the interior of the matter. This choice is not unique and other
continuations are possible, although they do not alter the values of the source
integrals, which we present below.

Using $W$ and $Z$ along $\mathcal S^{(2)}_0$ instead of $\rho$ and $\zeta$,
respectively, we can rewrite Eqs. \eqref{eq:multipole_moments} as surface
integrals:
\begin{align}\label{eq:surface_integrals}
 U^{(r)}=\frac{1}{8\pi}\int\limits_{\mathcal S^{(3)}_0}
     \frac{\e{U}}{W}\left(N_{-}^{(r)}U_{,\hat n}-N^{(r)}_{+,W}Z_{,\hat
     n}U+N^{(r)}_{+,Z}W_{,\hat n}U\right)\mathrm d \mathcal S_{0}^{(3)},
\end{align}
An integration by parts, the axially symmetry and the vacuum field equations are
necessary for this step. 

Using Stokes' theorem and the field equations we
obtain
\begin{align}\label{eq:volume_integral}
\begin{split}
 U^{(r)}=&\frac{1}{8\pi}
 \int\limits_{\mathcal
 V^{(3)}_0}\e{U}\left[-\frac{N^{(r)}_{-}}{W}
 R_{\alpha\beta}\frac{\xi^\alpha\xi^\beta}{\xi^\gamma\xi_\gamma} + 
 N^{(r)}_{+,Z}U \left(\frac{W^{,\alpha}}{W\phantom{{}^\alpha}}\right)_{;\alpha}
 -\right.\\
 &\left.  N^{(r)}_{+,W}U\left( \frac{Z^{,\alpha}}{W\phantom{{}^\alpha}}
 \right)_{;\alpha} +N^{(r)}_{+,WZ}\frac{U}{W}\big(W^{,\alpha}W_{,\alpha}-
 Z^{,\alpha}Z_{,\alpha}\big)\right]\mathrm d\mathcal V^{(3)}_0\\
 =&\frac{1}{8\pi}\sum\limits_{i}
 \int\limits_{\mathcal V_i^{(3)}}\e{U}\left[8\pi\frac{N^{(r)}_{-}}{W}
 (Tg_{\alpha\beta}-T_{\alpha\beta})\frac{\xi^\alpha\xi^\beta}{\xi^\gamma\xi_\gamma}
 + N^{(r)}_{+,Z}U \left(\frac{W^{,\alpha}}{W\phantom{{}^\alpha}}\right)_{;\alpha}
 -\right.\\
 &\left.  N^{(r)}_{+,W}U\left( \frac{Z^{,\alpha}}{W\phantom{{}^\alpha}}
 \right)_{;\alpha} +N^{(r)}_{+,WZ}\frac{U}{W}\big(W^{,\alpha}W_{,\alpha}-
 Z^{,\alpha}Z_{,\alpha}\big)\right]\mathrm d\mathcal V^{(3)}_i.
\end{split}
\end{align}
The $\mathrm d \mathcal V^{(3)}_i$ are the proper volume element of
$\mathcal V^{(3)}_i$ and a semicolon denotes the covariant derivative with
respect to the line element \eqref{eq:Weyl_line_element}. The last equality is
due to Einstein's equations, which imply that the integrand vanishes in vacuum.
The integrals \eqref{eq:volume_integral} are the desired source integrals. They determine the AMM from the geometry
inside the matter regions alone. Of course, Stokes' theorem can again be used to
rewrite the source integrals as surface integrals over $\mathcal S^{(3)}_i$ of the
respective matter component. In turn, these can be reformulated as line
integrals, cf. Sec. \ref{sec:Applications}. That the contributions of the
individual matter components $\mathcal V^{(3)}_i$ to the asymptotic multipole
moments superpose linearly is due to the choice of Weyl's multipole moments. If
we employ the method from \cite{Fodor_1989} to calculate Geroch's multipole
moments $M_r$ from Weyl's multipole moments $U^{(r)}$, we obtain a mixing
of the contributions $U^{(k)}_i$ of the individual matter components with $k<r$
in the $M_r$. This is already apparent for the quadrupole moment $M_2$, which
depends non-linearly on $U^{(0)}$:
\begin{align}
 M_2=U^{(2)}-\frac{1}{3}{U^{(0)}}^3.
\end{align}

The Geroch mass $M_0$ equals $U^{(0)}$ and is given by the (negative) Komar
integral.
This follows also from Eq. \eqref{eq:surface_integrals} with $r=0$.
% \begin{align}
%  M_0=\frac{1}{4\pi}\sum\limits_{i}
%  \int\limits_{\mathcal V^{(3)}_i}R_{ab}
%  \frac{\xi^a\xi^b}{\sqrt{-\xi^c\xi_c}}\mathrm d\mathcal V^{(3)}_i.
%  \end{align}

\section{Applications\label{sec:Applications}}

We conclude the paper discussing one possible application of the source
integrals \eqref{eq:surface_integrals}. Assume a matter distribution is given,
where the metric is known in the interior or the Dirichlet and the Neumann data
for $U$ are known at the surface. Even then it is far from trivial (at least in
the stationary case, see \cite{Ansorg_2002}) to obtain a global asymptotically
flat solution, if it exists. The source integrals for the AMM provide a tool to
solve this task. As a simple example serves here the case of static dust without
any surface distributions. In Weyl coordinates (not necessarily canonical) the
energy momentum tensor is given by
\begin{align}
T_{\alpha\beta}=\mu\e{2U}\delta_\alpha^t\delta_\beta^t.
\end{align}
The contracted Bianchi identities imply $U=\mathrm{const.}$ in the
interior and, thus, the gradient of $U$ vanishes at $\mathcal S^{(3)}_i$ in all
coordinates. Using the line integrals for Weyl's AMM, which follow from Eq.
\eqref{eq:volume_integral}, we get:
\begin{align}
 U^{(r)}=\frac{1}{4}\sum\limits_i\int\limits_{\mathcal S^{(2)}_i}
     \left(N_{-}^{(r)}U_{,\hat n}+N^{(r)}_{+} U_{,\hat s}\right)
     \mathrm d \mathcal S^{(2)}_i=0.
\end{align}
Thus, all AMM vanish and the spacetime is flat in the exterior. This contradicts
the presence of a dust distribution with positive mass density. Of course, this
result is already known and more general non-existence results for dust
including the rotating case can be found in \cite{Gurlebeck_2009,Pfister_2010}
and references therein. Although the non-existence is proved here, this example
shows in a concise way how the source integrals can be applied in more difficult
physical situations like rotating stars. This and other applications, e.g. to
tidal distortions of black holes, will be investigated in future work.

\section*{Acknowledgement} 
N.G. gratefully acknowledges support from the DFG
within the Research Training Group 1620 ``Models of Gravity'' and from the Grant
GACR-202/09/0772. The author thanks C. L\"ammerzahl, V. Perlick and O. Sv\'itek
for helpful discussions.

%\bibliographystyle{plain}
%\bibliography{References}
% 

\end{document}